\documentclass[amsmath,pre,epsfig,amssymb,notitlepage,superscriptaddress]{revtex4-1}
\usepackage[pdftex]{color,graphicx}
\usepackage{amsmath}
\usepackage{graphicx}
\begin{document}

\title{Quantitative study of crossregulation, noise and synchronization between microRNA targets in single cells}

\author{Carla Bosia}
\altaffiliation{Joint First Authors}
\affiliation{Human Genetics Foundation, Torino, Italy}

\author{Francesco Sgr\`o}
\altaffiliation{Joint First Authors}
\affiliation{Department of Applied Science and Technology, Politecnico di Torino, Italy}

\author{Laura Conti}
\affiliation{Molecular Biotechnology Center, University of Torino, Italy}

\author{Carlo Baldassi}
\affiliation{Human Genetics Foundation, Torino, Italy}
\affiliation{Department of Applied Science and Technology, Politecnico di Torino, Italy}

\author{Federica Cavallo}
\author{Ferdinando Di Cunto}
\author{Emilia Turco}
\affiliation{Molecular Biotechnology Center, University of Torino, Italy}

\author{Andrea Pagnani}
\altaffiliation{Joint Last Authors}
\affiliation{Human Genetics Foundation, Torino, Italy}
\affiliation{Department of Applied Science and Technology, Politecnico di Torino, Italy}
\author{Riccardo Zecchina}
\altaffiliation{Joint Last Authors}
\affiliation{Human Genetics Foundation, Torino, Italy}
\affiliation{Department of Applied Science and Technology, Politecnico di Torino, Italy}

\date{\today}


\begin{abstract}
{Recent studies reported complex post-transcriptional interplay among targets of a common pool 
  of microRNAs, a class of small non-coding downregulators of gene expression. 
  Behaving as microRNA-sponges, distinct RNA species may compete for binding to microRNAs and 
  coregulate each other in a dose-dependent manner. Although previous studies in cell populations 
  showed competition in vitro, the detailed dynamical aspects of this process, most importantly in
  physiological conditions, remains unclear. We address this point by monitoring protein expression
  of two targets of a common miRNA with quantitative single-cell measurements.
  In agreement with a detailed stochastic model of molecular titration, we observed that: (i) crosstalk between targets 
  is possible only in particular stoichiometric conditions, (ii) a trade-off on the number of microRNA 
  regulatory elements may induce the coexistence of two distinct cell populations, (iii) strong inter-targets
  correlations can be observed.
  This phenomenology is compatible with a small amount of mRNA target molecules per cell of the order of $10-10^2$.}
\end{abstract}

\maketitle



\section{Introduction}
Modern technologies to explore the transcriptome allow the
identification of many non-coding transcripts whose functions
are only partially known and that may control gene expression at different
levels. MicroRNAs (miRNAs) are small post-transcriptional repressors of gene
expression \cite{Flynt2008} belonging to this class of
molecules. Known to play crucial roles in several biological
processes, miRNAs often show altered expression profiles in human
diseases \cite{Bartel2004, Bushati2007, Alvarez-Garcia2005,
Esquela-Kerscher2006}. It is commonly believed that miRNAs play central roles
in conferring robustness to biological processes against environmental
fluctuations \cite{Li2009, Inui2010, Osella2011, Bosia12}. 
The common assumption that one miRNA molecule can at one time at most
interact with one target mRNA \cite{Marks2010} suggests a whole new layer of
post-transcriptional cross-regulation named the ``Competing Endogenous RNA (ceRNA) effect''
\cite{Salmena2011}. This theory proposes that the amount of a gene
product may be tuned by varying the concentration of another gene sharing with it
the same miRNAs. Qualitative experiments based on observing induced variations in transcripts 
indeed show that endogenous transcripts could be coupled due to the interaction with a common pool
of miRNAs \cite{Califano11, Tay11, Karreth11, Ala2013}.
The discovery that miRNA-target interaction is compatible with a titration mechanism
\cite{Mukherji2011} supports the emergence of hypersensitivity regions
\cite{Elf2003, Buchler2008} where miRNA targets should be highly
correlated and their relative stoichiometry tightly controlled
\cite{Bosia2013, Riba2014}. However, the relevance of the ceRNA effect is
still largely debated: while absolute quantification experiments in
primary hepatocytes and liver cells suggests that the ceRNA effect is unlikely
to significantly affect gene expression and metabolism
\cite{Bartel14}, differential susceptibility based on endogenous
miRNA/target pool ratios provided a physiological context for
ceRNA competition in vivo \cite{Sharp14}. Crosstalk among mRNAs may
thus be regulated depending on miRNA and mRNA relative abundances and
may exhibit a complex phenomenology in terms of target correlation
and relative fluctuation profiles \cite{Bosia2013}.

Here, we experimentally explore these features
addressing the relevance of the relative mRNA-miRNA stoichiometric composition.
Through the design of two bidirectional plasmids, each with a two-color fluorescent reporter system,
we simultaneously tracked gene expression in the presence and absence of miRNA regulatory elements (MRE). 
A stochastic gene interaction model \cite{Bosia2013} predicts the parameter region in
which the targets are most correlated. 
Using flow cytometry measurements of cotransfected mammalian cells allowed us to quantify the
crosstalk and correlations predicted by the model. 
We found that, besides the well-known ``sponge effect'' a given transfected target
can exert on the others \cite{Ebert2010}, there is an optimal range of parameters (in
terms of effective transcription rates and miRNA interaction strengths) for which crosstalk is possible among ceRNAs. 
We show that such regulation arises both at the level of mean protein concentrations
and noise and that it is compatible with low numbers of mRNA molecules. 
It is worth noting that an absolute quantification of exogenous
transcripts reveals that in our experiments the crosstalk is highest in a {\it physiological} 
regime of order $10$ to $10^2$ molecules per cell \cite{Selbach2011, Marinov2014}.
Moreover, there is a non-trivial competition mechanism on the number of available MRE such that synchronization
can arise together with low noise. 
Interestingly, in agreement with the model, the same mechanism may induce bimodal population distributions 
with distinct high and low expression states of the targets.

\section{Stochastic titration model for crosstalk}

We suggest a stochastic model for the miRNA-mediated target
crosstalk that provides insight into target cross-regulation 
\cite{Levine2007, Mukherji2011, Bosia2013}{} (see Figure 1a). Through the
formulation of a chemical master equation (see Material and Methods and Supplementary Information (SI) for
details on the model), the model describes the
amount of two free mRNAs $r_1$ and $r_2$ which are both targets 
of the same miRNA $s$, as a function of their constitutive expression $r_0$ (i.e. the value
of $r_1$ or $r_2$ when $g_1$ or $g_2$ tend to $0$).
$r_1$ and $r_2$ can be translated into proteins ($p_1$ and $p_2$ respectively). 
The two candidate ceRNAs $r_1$ and $r_2$, and thus $p_1$ and $p_2$, are
coupled through their common miRNA $s$ which can bind both of them and
then be released with or without degradation of $r_1$ or $r_2$ (miRNA
turnover). The pool of available mature miRNAs is then the
limiting factor in a system of potentially interacting targets. 
A Gaussian approximation of the master equation allows us to evaluate mean
values and to quantify noise (coefficient of variation $CV_x = \sigma_{p_x}/p_x$) for
$p_1$, $p_2$ and $p_0$ (the protein translated from $r_0$) as well as the 
Pearson correlation coefficient between $p_1$ and $p_2$ (i.e. $(\langle p_1 p_2 \rangle - \langle p_1
\rangle \langle p_2 \rangle)/\sigma_{p_1}\sigma_{p_2}$), see Figures 1c-e respectively. 
Two parameters ($g_1$ and $g_2$) determine qualitatively the shapes of the
functions generated by the model. These parameters are proportional to the miRNA-mRNA association rate. 
When one of them tends to zero (say $g_2$) then its corresponding target
($r_2$) is not interacting with the miRNA, while the other ($r_1$)
behaves as described by Mukherji and coworkers \cite{Mukherji2011}{}: as $g_1$ increases, $r_1$ (and then $p_1$) is
repressed until a threshold level of $r_0$ is exceeded (Figure 1c). 
The threshold established by miRNA regulation and the increase of $g_1$
sharpens the transition between threshold and escape regimes. From the
point of view of $r_1$, $g_2$ (proportional to the association
constant of the second target) governs the concentration of free miRNA
available within the cell. Increasing $g_2$ (keeping all
other parameters fixed) pushes the threshold to lower values of expression 
(lower $r_0$) and globally increases $r_1$ (and $p_1$): $r_2$ is behaving as a sponge
for the miRNA, and increasing $g_2$ is equivalent to sponge away the
miRNA available to target $r_1$. When all the miRNA has been sponged
away by $r_2$ (high value of $g_2$), then $r_1$ is not regulated anymore 
and its mean value is simply $k_{r_1}/g_{r_1}$
(with $k_{r_1}$ and $g_{r_1}$ transcription and degradation rates of
$r_1$ respectively). In an intermediate situation in which miRNA is
not completely sponged away by one of the targets, finely-tuned
crosstalk between targets is possible.  The mathematical model thus
suggests experiments to perform in order to test this hypothesis and
to quantify the crosstalk, modulated by $g_1$, $g_2$ and the amount of
miRNA present in the cell.

\section{Flow cytometry reveals cross regulation}

To investigate the predicted miRNA-mediated crosstalk in single mammalian cells, we
used two different two-color fluorescent reporters, as sketched in
Figure 1b. Both constructs consist of bidirectional promoters driving
two genes whose products are fluorescent proteins. The first
construct expresses the fluorescent proteins mCherry and enhanced
yellow fluorescent protein (eYFP) \cite{Mukherji2011}, while the second contruct 
expresses mCerulean and mKOrange. The 3'untranslated region (UTR) of both mCherry and
mCerulean was engineered to contain a fixed number $N$ of MRE for miR-20a
(with $N=0,1,4,7$), a miRNA endogenously expressed by HEK 293 cell
line \cite{Bartel2009, Ruegger12}. mCherry and mCerulean are therefore proxies for
the two targets in the model. The 3'UTRs of eYFP and mKOrange were left unchanged in order to
measure the transcriptional activity of the reporters in single
cells. The constructs thus allow simultaneous monitoring of protein
levels with (mCherry and mCerulean) and without (eYFP and mKOrange) miRNA regulation. 
In the case of single construct transfections, when individual cells are sorted according to
their eYFP or mKOrange levels, we observed the threshold effect
documented for HeLa cells by Mukherji and coworkers \cite{Mukherji2011}. 
Briefly, when no MRE are present, mCherry (mCerulean) and eYFP (mKOrange)
levels of expression are proportional. In cells with one or more
miR-20a sites on mCherry (mCerulean), the mCherry (mCerulean) level
does not increase until a threshold level of eYFP (mKOrange) is
exceeded (see Figure 5). This indicates that the protein production is highly repressed 
below the threshold established by miRNA regulation and responds sensitively to target mRNA input close to it.
Cotransfections of both constructs with different MRE numbers
and measurements of fluorescence with flow cytometer enabled the
quantification of crosstalk between mCherry and mCerulean as a
function of $N$. We expect both mCherry and mCerulean
signals to be proportional to $p_1$ and $p_2$ respectively, while eYFP (mKOrange)
is proportional to $p_0$. To quantitatively capture the crosstalk, we measured the joint
distributions of mCherry ($p_1$) and eYFP ($p_0$) levels given mCerulean ($p_2$) in single
cells positive to the fluorophores. We then binned the data according
to their eYFP levels and calculated the mCherry and mCerulean mean
levels as well as standard deviations in each eYFP bin.
Transient cotransfections allowed us to
explore the space of parameters and manage $10^5-10^6$ cells.
Our analysis has been restricted to cells whose fluorescence was 95\%
confidence above cellular autofluorescence and is independent of the transfection method.

\section{Phenomenology of crosstalk} 

To quantify the crosstalk by modulating $g_1$ and $g_2$, we performed
cotransfections with different MRE on mCherry and mCerulean ($N
= 1,4,7$) and compared them with the case $N=0$. We thus obtained 16
combinations of different cotransfections which allowed us to follow
the expression of one target (mCherry) while tuning the amount of free miRNA
via the second target (mCerulean). Our results are independent of the method of
transfections as similar results were obtained by transfecting cells with lipofectamine-based 
(data reported in Figures 2-5,7,8) or CaCl$_2$-based protocols (Figure 6).
As predicted by the model (Figure 1c), it is possible to identify two different effects: (i) the
appearence of a threshold on mCherry while increasing the number $N$ of MRE on its 3'UTR
and keeping $N=0$ on mCerulean (Figure 2a and Figure 6a) and (ii) a global increase
of mCherry mean fluorescence and a shift in the threshold while
increasing $N$ on mCerulean (Figure 2b and Figure 6b). mCherry thus tends to the 
unregulated case (mCherry linearly proportional to eYFP) while increasing the number of MRE on mCerulean. 
This result is well summarized by the fold repression $F$ between
regulated and unregulated mCherry mean fluorescence (Figure 2c-e). 
$F$ is the ratio between the value of mCherry in the absence of miR-20a 
MRE and its value in the presence of MRE for each eYFP bin and for each $N$ on mCerulean.
Increasing the number of MRE on mCherry increases its repression, and $F$ is highest when mCerulean 
has $N=0$ MRE while tends to one increasing eYFP or the number of MRE on mCerulean. 
In particular, in proximity to the threshold, $F$ shows a maximum whose value depends both on mCherry and mCerulean MRE.
$F$ could be indirectly considered as a measure of crosstalk between the two targets. 
Crosstalk is maximal for intermediate levels of repression, when mCerulean has between 1 and 4 MRE. 

\section{Local increase of cell-to-cell variability: bimodality}

It is well known that the intrinsic noise of an unregulated gene product decreases
when its expression level increases \cite{Kaern05}. 
The effect of miRNA regulation could introduce an extra source of
noise (extrinsic noise). Our mathematical model predicts that, at fixed levels of
expression, the total noise (intrinsic plus extrinsic) of a miRNA-regulated gene product should increase
upon enhancing miRNA-target interaction strength (see Figure 1d) with
respect to the unregulated case.
In particular the model predicts the onset of a local maximum in the noise profile of a miRNA target
versus its level of constitutive expression for high miRNA-target
interaction strength. Experimentally, we could identify
two competing effects: (i) upon increasing $N$ on mCherry (i.e. $g_1$)
the total noise of mCherry, quantified by its CV, globally increases as a function of eYFP (Figure 3a and Figure 6c) and
(ii) upon increasing $N$ on mCerulean (i.e. $g_2$) the total noise of mCherry globally decreases 
(Figure 3b and Figure 6d). The overall result is that there is an optimal
range of MRE on the ``sponge'' (mCerulean) for which a given
miRNA-regulated target (mCherry) can show lower noise with respect
to a target with lower $N$ (compare Figures 3a and 3b). For high levels of repression
(high $N$ on mCherry and low $N$ on mCerulean), mCherry CV eventually shows a local
maximum in proximity to the threshold (Figure 6c,d). 
A low level of noise indicates unimodal distributions while an increase
in noise stands for an increased cell-to-cell variability and may indicate bimodal population distributions 
with distinct high and low expression states \cite{Blake03}.
We then checked if this was the case and found that bimodality
on mCherry is present near the threshold in case of high miRNA-target interaction
($N=4,7$ on mCherry and $N=0,1$ on mCerulean), see histograms in Figure 3 and Figure 7. 
In particular, for $N=7$ on mCherry and $N=0$ on mCerulean two well discernible phenotypes appear.
This suggests the binary response is directly linked to the
variability in the level of repression the miRNA exerts on the
target. The emergence of bimodality in the proximity of the miRNA-target threshold has been recently suggested \cite{Bosia2013}.


\section{Shift of the optimal crosstalk region} 
In order to assess the crosstalk dependence on the availability of miRNA 
we transfected $100 nM$ of pre-miR for miR-20a together with the bidirectional constructs. 
In our model this is equivalent to increasing the basal miRNA transcription rate $k_s$. 
We analyzed the cases with $N=4$ for mCherry and $N=0,1,4,7$ for
mCerulean. In agreement with the model predictions and with previous work \cite{Mukherji2011}, 
we observed a shift of the threshold towards
higher eYFP levels (Figure 4a) together with a global increase
in the fold-repression (Figure 4b) and a resulting shift of the optimal crosstalk
region towards a higher number of MRE. We then
quantified the absolute amount of exogenous targets in three
subpopulations of cells, sorted according to their eYFP intensity
(low, medium and high) both in the presence and absence of pre-miR for the
case with $N=4$ on mCherry and $N=1$ on mCerulean (Figure 4c). We found that
mCherry and mCerulean ranged from 40 to 400 and from 10 to 250
molecules per cell, respectively, without pre-miR and both from 10 to about 100
molecules in the presence of pre-miR. In particular, the
amount of exogenous targets for an intermediate level of eYFP is in
quantitative agreement to what previously found \cite{Mukherji2011}. 
These values show that even if the phenomenology so far described has been
obtained through transient cotransfection, the numbers involved could be
compatible with physiological values \cite{Sharp14}.

\section{Potential synchronization in permissive environment} 
Our model predicts a maximum in the correlation between two 
miRNA targets near the threshold (see Figure 1e). 
We investigated the strength of this prediction distinguishing between correlations
dependent on the experimental setting (mainly transient cotransfections and partial sharing of regulatory
elements in the promoter) and correlations induced by the competition for miRNA binding,
which can potentially lead to synchronized fluctuations. 
We thus defined the ratio of the Pearson correlation coefficients (ratio of Pearson coefficient
between mCherry and mCerulean possessing different MRE to the same measure in the absence of MRE). 
We measured this ratio for eYFP below, around and above the threshold (Figure 8a-c respectively),
and observed that the competition for miRNA binding introduces correlations ranging from 4 to 12 fold 
higher than the basal level of correlation. 
Our results show that it is possible to have weakly or highly correlated targets for precise transcriptional programs. 
The regime of synchronized fluctuations is determined by the number of MRE on both the targets and by their relative stoichiometry. 

\section{Discussion}
Our results on miRNA-mediated target cross-regulation offer a detailed feature map to characterize the ``ceRNA effect'' (Figure 9).
Besides the general consistency with previous population-based qualitative results \cite{Ala2013}, and the 
agreement with a titration-based mechanism of miRNA-target interaction \cite{Mukherji2011, Bosia2013, Figliuzzi13}, the stochastic analysis allowed us to characterize curve trends for fluctuations and correlations of two miRNA targets as function of their expression level.
The detailed picture points out that crosstalk between targets is quantitatively relevant only in conditions of intermediate miRNA repression
and small amounts of target molecules (order $10-10^{2}$), in agreement with a cell population-based study by Bosson and coworkers \cite{Sharp14}. Since in this situation two or more targets may be highly cross-correlated, 
our result suggests that optimal levels of expression of genes and of miRNAs with respect to maximizing crosstalk may control relative fluctuations of targets that have to interact or bind in complexes with a precise stoichiometry \cite{Riba2014}.
On the other hand, we found that strong miRNA repression together with low target crosstalk is sufficient to induce bimodality, i.e. the appearance of two distinct population of cells with low and high target expression states. 
This result suggests that the system could be locked in one of these two states both changing the miRNA-target interaction strength through the expression of other competitors or through regulatory links. 
Such titrative regulatory mechanisms of miRNAs may easily switch `on' or `off' whole gene networks depending on miRNAs and targets relative stoichiometry. It is thus tempting to speculate that gene expression thresholding could be an important feature of cell fate decisions.

Although our experimental setting is ``artificial", i.e. induced by transient cotransfections of engeneered plasmids, it provides a deep exploration of the parameter space. A physiological system of miRNAs and targets could indeed experience only a small subset of the features so far described, rendering the characterization of crosstalk difficult. 
The map of properties we ended up with (Figure 9) links quantitative measurements (effective miRNA repression and number of mRNA molecules) and model parameters (effective miRNA-target binding and transcription rates), suggesting the possibility to move around in phenotype space tuning quantities as the accessibility of binding sites or the affinity between miRNA and targets. In particular, it suggests the class of molecules we should look at when investigating for crosstalk. 
Molecular species physiologically present in the order of $10-10^2$ molecules per cell, such as transcription factors or signalling molecules \cite{Marinov2014,Weissman2014}, are more likely to be affected by crossregulation since they are potentially closer to the threshold than highly expressed genes.
We think that this findings would pave the way for signifincant progress in the understanding of pathological mechanisms underlying many 
diseases based on miRNA-target crosstalk dysregulation and therefore potentially lead to new therapeutic options.\\
During the submission process of this paper we became aware of ref.\cite{Wang2015} which contains results that partially overlap with a subset of ours.

\section{Materials and methods}
A detailed description of experimental procedures used in this study
(including reporter plasmid construction, cell transfection, FACS
measurement, cell sorting, qTR-PCR, modelling and data analysis
procedures) is available in SI.

\section*{Authors Contribution}
Carla Bosia (CB1), AP and RZ derived the stochastic model; CB1,
FDC, ET, AP and RZ designed experiments; CB1, Carlo Baldassi (CB2) and
AP analyzed data; CB1, FS, LC, FC and ET performed experiments; CB1,
FS, LC, FDC, AP and RZ wrote the manuscript.
\section*{Corresponding Author}
Carla Bosia: {\tt carla.bosia@hugef-torino.org}

\begin{acknowledgments}
CB2 and RZ aknowledge the European Research Council for grant
n. 267915. The authors would like to thank E. Fraenkel, A. Weisse and
M. Osella for carefully reading the draft and U. Ala, F. Balzac,
M. Caselle, P.P. Pandolfi, P. Provero, M. Osella and A. Riba for
helpful discussions.
\end{acknowledgments}



\begin{figure}
\centerline{\includegraphics[width=0.5\columnwidth]{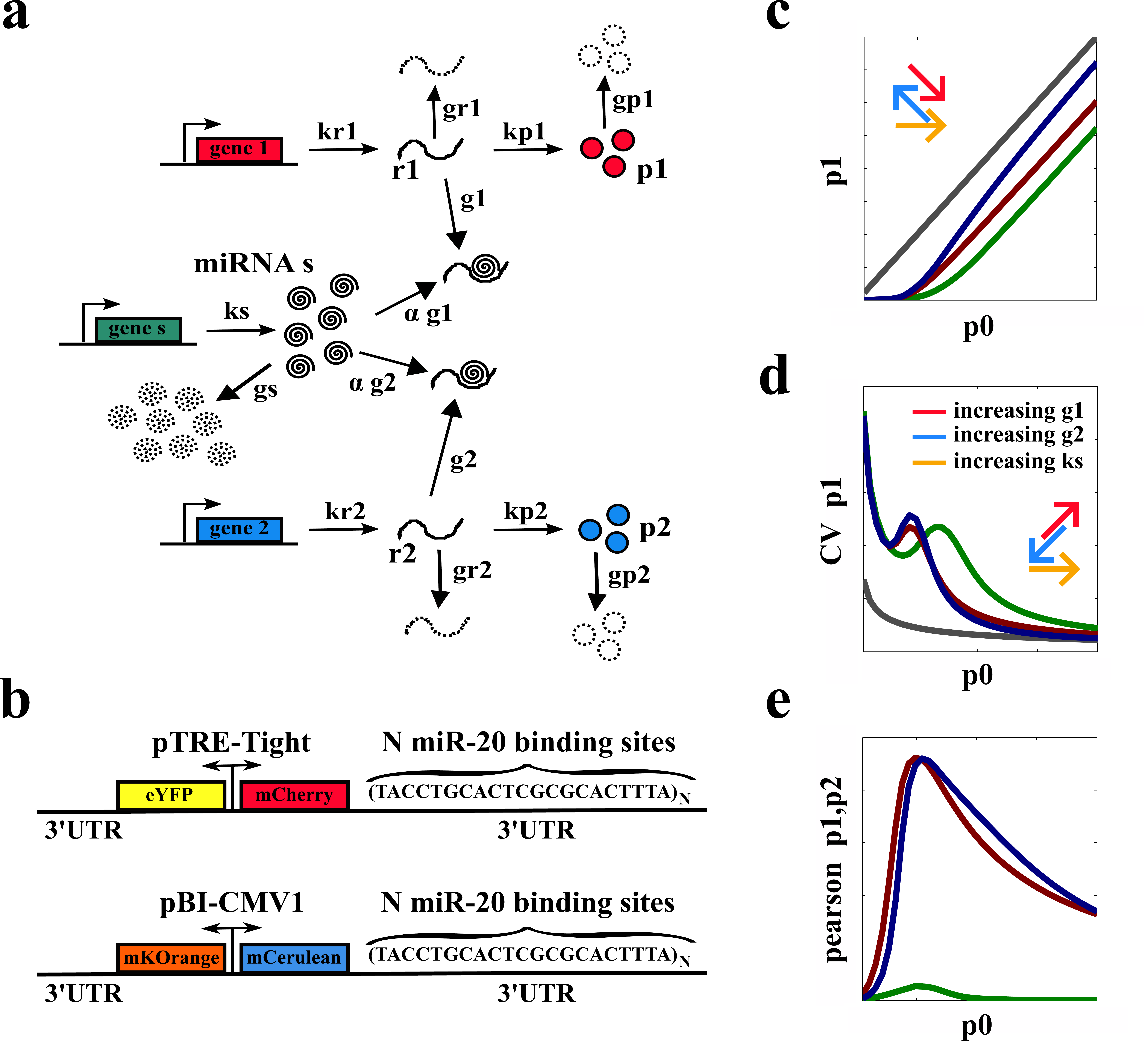}}\protect\caption{\label{fig:1}\textbf{Model and predictions}
(a) Sketch of the minimal model of miRNA-target interactions. One miRNA $s$ and two targets $r_1$ and $r_2$ are independently transcribed 
with rates $k_s,k_{r_1}$ and $k_{r_2}$, respectively. Each transcript can then degrade with rate $g_s,g_{r_1}$ and $g_{r_2}$, respectively.
Each miRNA $s$ can interact with targets $r_1$ or $r_2$ with an effective binding rates $g_1$ or $g_2$. $\alpha$ measures the probability of miRNA recycling. If not bound to a miRNA, targets $r_1$ and $r_2$ can be translated into proteins $p_1$ and $p_2$ respectively, which could then degrade with rates $g_{p_1}$ and $g_{p_2}$.
(b) Schematic representation of the two bidirectional plasmids coding for
the four fluorophores. (c-e) Predictions from the stochastic model of interactions sketched in (a) in terms of mean amount of $p_1$ free molecules (c), $p_1$ coefficient of variation $CV_{p_1}$ (d) and pearson correlation coefficient between $p_1$ and $p_2$ (e) as a function of $p_0$, which is the constitutive value of $p_1$ when $g_1$ tends to $0$. The colored arrows in (c) and (d) show the directions in which the curves move when tuning the miRNA-target interaction strengths $g_1$ and $g_2$ and the miRNA transcription rate $k_s$. Grey curves in (c,d) are
predictions for $p_1$ and $CV_{p_1}$ when $g_1$ tends to $0$.}
\end{figure}

\begin{figure}
\centerline{\includegraphics[width=0.5\columnwidth]{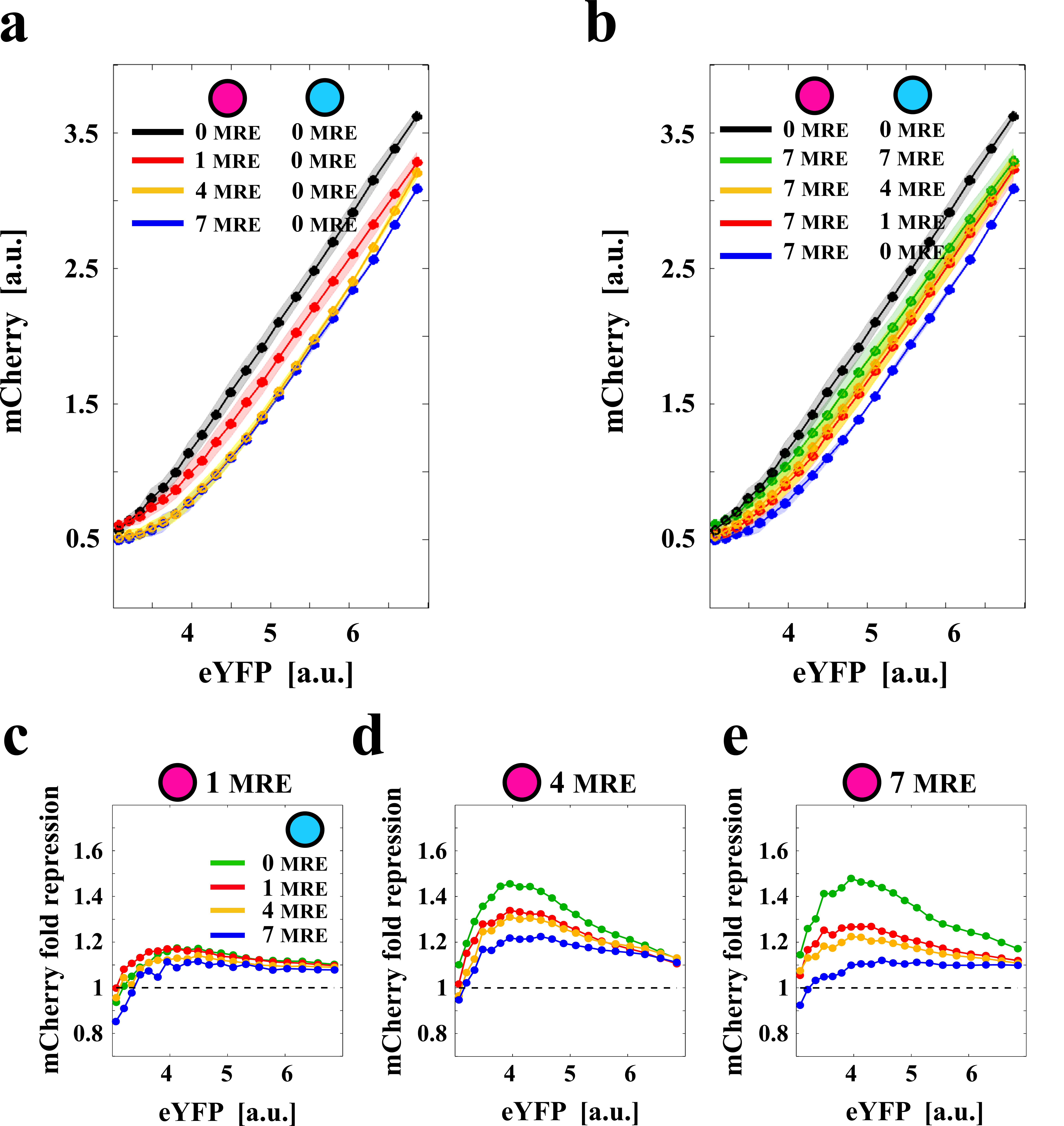}}\protect\caption{\label{fig:2}\textbf{Threshold and fold-repression}
  (a-b) mCherry mean fluorescence is plotted against eYFP. Shadowed strips around data are the error bars on the biological replicates. A threshold emerges when increasing mCherry MRE (a) while it disappears when increasing mCerulean MRE (b). The intensity of crosstalk (measured in terms of fold-repression $F$ with respect to the unregulated fluorophores) depends on the particular combination of MRE on both exogenous targets (c-e). Purple and cyan circles in legends represent the plasmids coding for the mCherry and mCerulean fluorophores.}
\end{figure}

\begin{figure}
\centerline{\includegraphics[width=0.5\columnwidth]{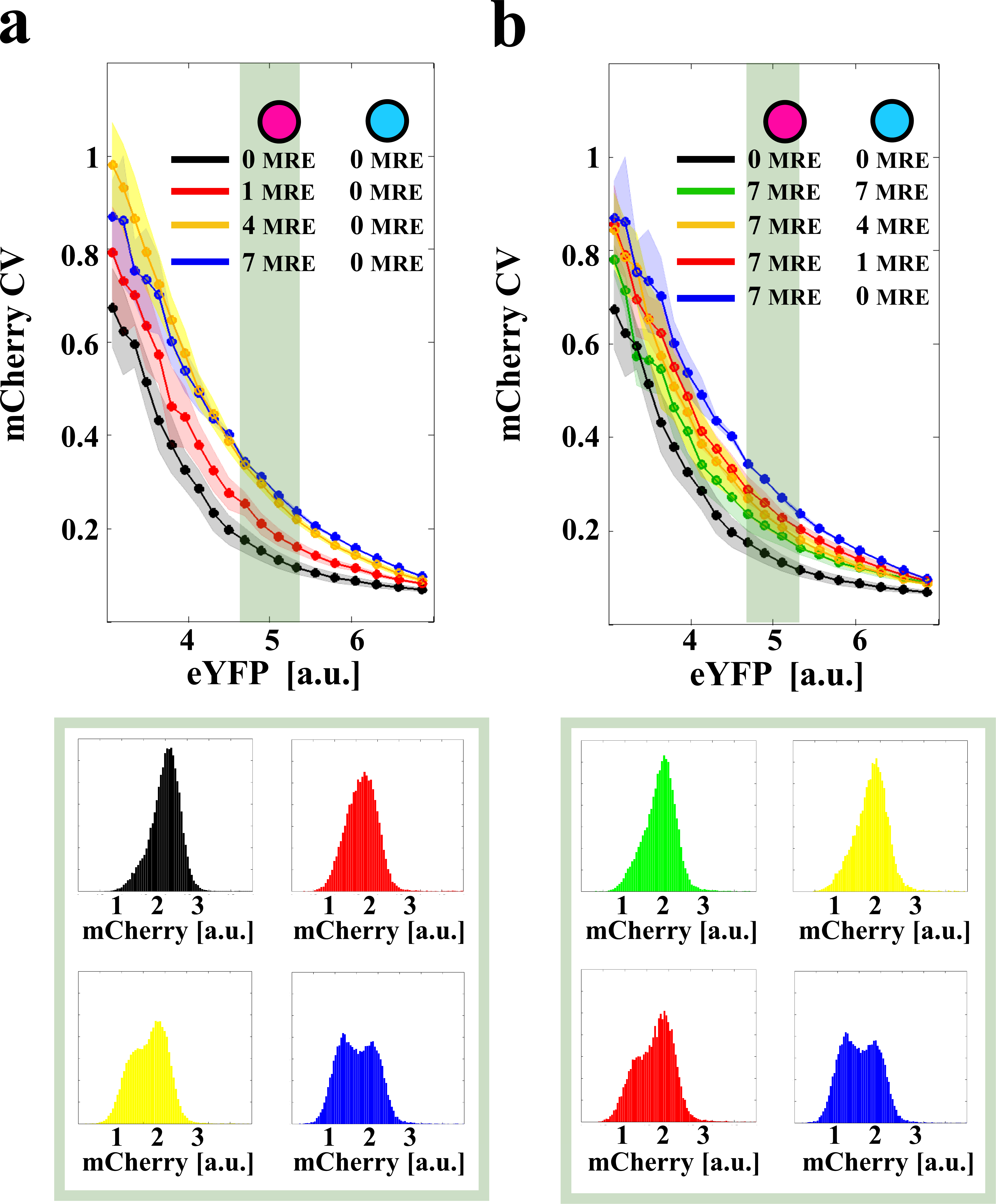}}\protect\caption{\label{fig:3}\textbf{Local increase of cell-to-cell variability}
  (a-b) mCherry total noise quantified by its coefficient of variation (CV) is plotted against eYFP. Shadowed strips around data are the error bars on the biological replicates. The CV globally increases when increasing the number of mCherry MRE (a) while decreases when increasing mCerulean MRE number (b). The competition of these two ``strengths” has the result of lowering the noise even if the expected repression from the rough number of mCherry MRE is high. Histograms in the lower panels show mCherry data distributions for the shaded regions in (a-b). A strong miRNA target repression strength increases cell-to-cell variability with the eventual appearance of different phenotypes (bimodal distributions). Purple and cyan circles in legends represent the plasmids coding for mCherry and mCerulean fluorophores, respectively.}
\end{figure}

\begin{figure}
\centerline{\includegraphics[width=0.7\columnwidth]{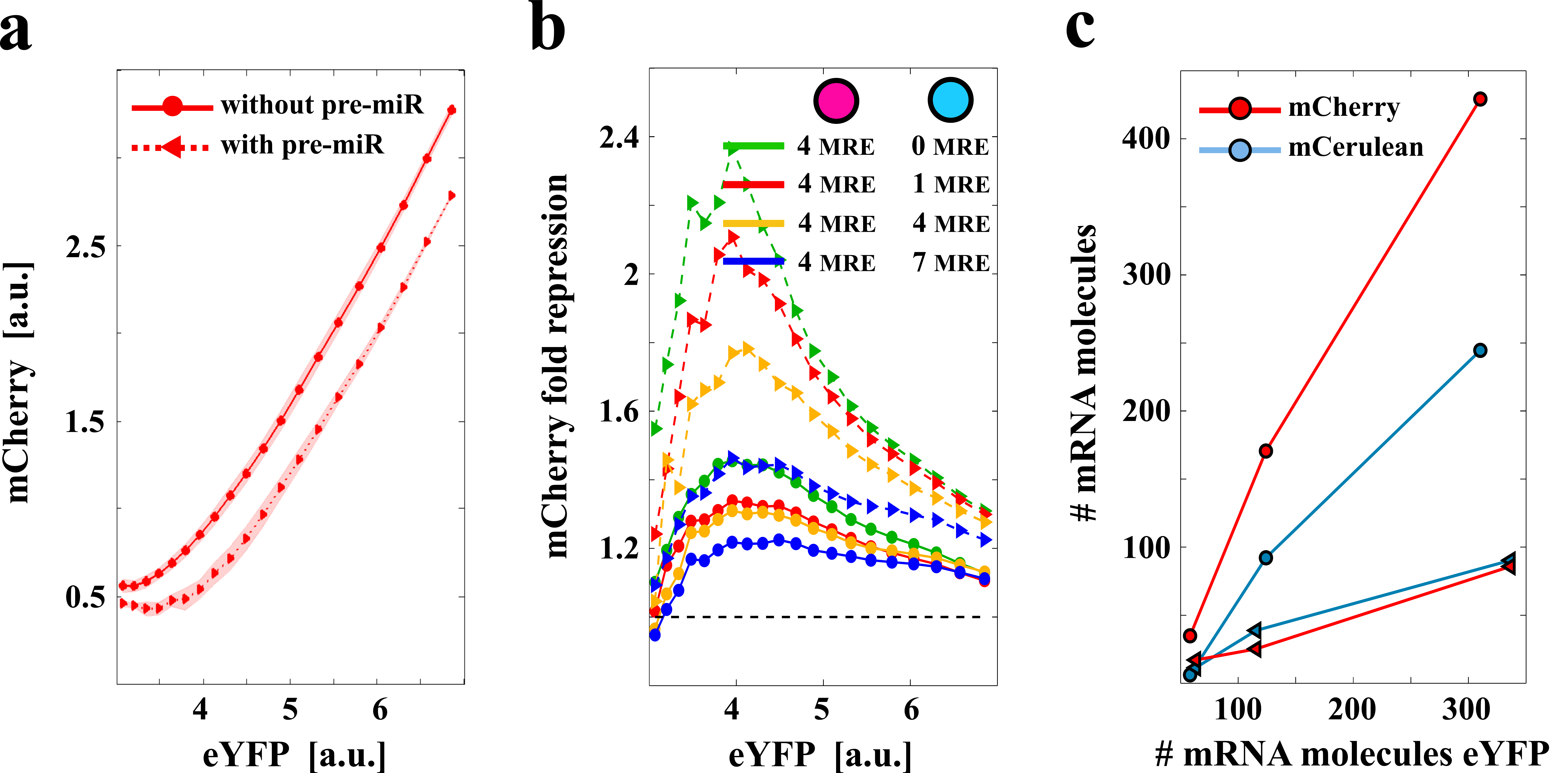}}\protect\caption{\label{fig:4}\textbf{Shift of the optimal crosstalk region}
  (a) According to the model, increasing the pool of available miRNAs (transfecting pre-miRNAs) shifts the threshold to higher constitutive expression values. Shadowed strips around data are the error bars on the biological replicates. (b) Different combinations of miR-20a MREs lead to different levels of fold repression and crosstalk. Purple and cyan circles in legends represent the plasmids coding for mCherry and mCerulean fluorophores, respectively. Triangles and circles in the plot are data from transfections with pre-miR20a and negative control respectively. (c) The mean amount of mRNA exogenous molecules per cell for three different intervals of eYFP basal expression is low enough to be comparable with physiological values. Triangles and circles show quantification in presence and absence of pre-miR-20a respectively.}
\end{figure}

\clearpage

\section*{Supplementary Information}
\subsection{Reporter plasmid construction}
The set of fluorescent reporters coding for eYFP and mCherry was
obtained from Addgene ($\#31463, \#31464, \#31465, \#31466$, deposited
by Phil Sharp Lab) and are the same used in \cite{Mukherji2011}{}.  
The second set of fluorescent reporters were cloned into pBI-CMV1
(Clontech). NLS sequence (ATGGGCCCTAAAAAGAAGCGTAAAGTC) was appended to
mCerulean-N1 (Addgene $\#27795$, deposited by Steven Vogel Lab
\cite{Koushik06}) by PCR and then inserted into the main vector with
ClaI and BamHI. mKOrange-NLS (Addgene $\#37346$, deposited by Connie
Cepko Lab \cite{Beier11}) was cloned into the vector using EcoRI blunt
and BamHI. miR-20a regulatory elements were appended to the 3'UTR of mCerulean with the same 
strategy applied in \cite{Mukherji2011}{}

\subsection{Transient transfections}
We performed two different methods of transfection, with Lipofectamine (data in
Figures 2-4 in the main text) and with CaCl$_2$ (data in Figures 1S-3S in SI).
Lipofectamine transfection method: 293-HEK TeT-Off cells (Clontech) below passage 6 were plated in G418
(Gibco) $200 \mu g/ml$ media in 6-well dishes the day before transfection. 
Reporter plasmids were transfected with Lipofectamine 2000 (Invitrogen) following the
manufacturer's specifications. miR-20a pre and negative control (Ambion) were
cotransfected at the indicated concentrations. Media was changed 24 h
after transfection. Assays were performed 48 h after transfection.
CaCl$_2$ transfection method: 293-HEK TeT-Off cells (Clontech) below passage 6 were plated 
in $100\times20$ mm (Falcon BD) dishes the day before transfection.
The cells were transfected using CaCl$_2$ protocols \cite{Jordan1996}{}.
Media was changed 24 h after transfection. Assays were performed 48 h after transfection.

\subsection{Flow cytometry}
Cells were harvested 48 h after transfection (cell confluency
$\sim90\%$) and run on a CyanADP (Beckman Coulter) flow cytometer. For
each sample, at least $0.5 \cdot 10^6$ cells were acquired. The raw FACS
data were analyzed with Summit3.1 software (Beckman Coulter) to gate
cells according to their forward and side scatter profiles and to
define the intensity of fluorescent signals emitted by the four
reporters in each cell. These values were normalized for background
fluorescence by subtracting the mean plus two standard deviation of
the fluorescent signal measured in the unstransfected control
cells. Data were then binned according to their eYFP values.

\subsection{Fluorescence-activated cell sorting}
Cells were transfected with the $N=4$ eYFP-mCherry and $N=1$ 
mkOrange-mCerulean reporters and pre-miR-20a 100 nM (Ambion; PM10057)
or Negative miRn20 100nM (Ambion). 48 h after transfection three cell
populations were sorted according to their eYFP fluorescence value 
(low, medium and high YFP expression) using a BD FACS Aria III (Becton Dickinson) 
cell sorter. Cell pellets were washed and snap frozen before RNA isolation.

\subsection{Quantitative Real-Time Polymerase Chain Reaction}
Total RNA was extracted using Trizol reagent (Ambion Life Technologies, USA) 
in combination with Pure Link RNA Mini Kit (Ambion) from each sorted cell subpopulation. $1 \mu g$ of total RNA 
was reverse transcribed using M-MLV reverse transcriptase and random primers (Life Technologies). 
Quantitative rt-PCR was performed on a 7300 Real Time PCR System (Applied Biosystems) using specific primers for eYFP, mCherry,
mCerulean and mKOrange. 18S probe (Life Technologies) was used as
internal control. Subsequent diluitions of each amplicon in known volumes allowed the definition of a calibration curve for each fluorophore
directly linking threshold cycles and number of molecules per cell.

\subsection{Empirical observables and Pearson correlation coefficient ratio}
We defined the empirical average of a given observable $O$ over an ensamble of cells with the symbol $\langle O \rangle = \sum_{i\in\mathrm{cell}} O_i/N_{\mathrm{cell}}$.
The Pearson ratio is defined as the ratio of the Pearson correlation coefficient
($\rho_{x,y} = (\langle xy\rangle-\langle x\rangle \langle y\rangle)/\sigma_x \sigma_y$) between mCherry
and mCerulean with a given combination of MRE to the same measure in absence of MRE.
We evaluated the ratio for each eYFP bin (below, around and above threshold) for at least three 
different biological replicates. We then estimated the p-values of each ratio with respect to the 
distributions having as standard deviation the error on biological replicates and as mean values the 
pearson ratio for mCherry and mCerulean with $N=0$ MRE for the three eYFP intervals.

\subsection{Stochastic model of molecular titration and crosstalk}

\subsubsection{Model definition}

We describe with a stochastic model the miRNA-target interactions. 
The system can be described by 5 interacting variables (1 microRNA, 2 mRNAs, 2 proteins) indicated respectively as $s, r_1, r_2, p_1,
p_2$, which represent the (integer) copy number of molecules present in the cell at any given time $t$. Using this notation, the
probability $P$ of finding in a cell exactly $s, r_1, r_2, p_1, p_2$ molecules at any time $t$ is governed by the following
master equation:
\begin{eqnarray}
\label{eq:master}
\partial_t P &=& 
\sum_{i=1}^2 \left[ k_{r_i}(P_{r_i-1} - P) + k_{p_i} r_i (P_{p_i-1} -
P) \right] + 
k_s(P_{s-1} - P)\nonumber\\
&+& 
\sum_{i=1}^2 \left\{ g_{r_i} [ (r_i + 1) P_{r_i+1}-r_iP ] + 
 g_{p_i} [ (p_i + 1) P_{p_i+1}-p_iP ] \right \} + 
            g_s    [ (s  + 1) P_{s+1}-sP]\nonumber\\
&+&
a \sum_{i=1}^2 g_i [(r_i+1)(s+1) P_{r_i+1,s+1} - r_i s
       P]\nonumber \nonumber\\
&+&
(1-\alpha)\sum_{i=1}^2 g_i s [(r_i+1) P_{r_i+1} - r_i P]
\end{eqnarray}
where $P:= P_{r_1,r_2,p_1,p_2,s}$ and, for example $P_{p_2+ 1}$ is a
short hand notation for $p_{r_1,r_2,p_1,p_2 + 1,s}$. In
Eq.~(\ref{eq:master}) $k_{r_i}, k_s, k_{p_i}$ $i=1,2$, are the
transcription rates of mRNAs $r_i$ and microRNA $s$ and the translation rates for proteins $p_i$
respectively. $g_{r_i},g_{p_i},g_s$ $i=1,2$ are their degradation
rates. $g_{i}$ $i=1,2$ are the effective association rates for the
microRNA $s$ and the mRNA $r_i$. Finally the parameter $\alpha$ measures
the catalyticity of the interaction, {\em i.e.} the fraction of
microRNA molecules that are recycled after the interaction with their
targets. This master equation is not amenable for analytic solutions and approximate methods have been proposed
\cite{Ala2013, Bosia2013} to obtain accurate quantitative predictions.
Following previous work \cite{Bosia2013} we obtained the approximated
expression for mean values, standard deviations and Pearson correlation
coefficients (sketched in Figure 1c-e), which we describe in the next paragraph.

\subsubsection{Independent molecular-species approximation}

As long as one is interested in mean values of the observables at
steady state, a good approximation is the so-called independent
molecular species approximation, also known as mean-field
approximation which amounts to assume that the multivariate
probability distribution $P$ is factorized among the different
chemical species:
\begin{equation}
\label{eq:factorized}
P^{\mathrm{ind}}(r_1,r_2,p_1,p_2,s) := p_{r_1}(r_1)
p_{r_2}(r_2) p_{p_1}(p_1)  p_{p_2}(p_1)  p_{s}(s)\,\,\,.
\end{equation}
Plugging this factorized functional form into Eq.~(\ref{eq:master}),
and computing the first moments at steady state ({\em i.e.} in the limit
$t\rightarrow \infty$), one obtains a system of second order equations
in the five variables which is easily solved numerically for any value
of the model parameters. The main limitation of the factorized ansatz
in Eq.~(\ref{eq:factorized}) is that, although it empirically turns
out to give a fairly accurate prediction of the mean values of the
different chemical species across a wide range of parameters, the very
simple structural form of Eq.~(\ref{eq:factorized}) cannot predict
their statistical correlations such as Pearson correlation
coefficients which under the independent chemical species
approximation are always zero by definition.

\subsubsection{Gaussian Approximation}

To overcome the above mentioned limitations and to take under control
correlations across chemical species, a very simple yet accurate
approximation scheme is the so-called Gaussian one
\cite{Bosia2013}. Note that, following this approximation scheme copy
numbers will not be bound to be integer numbers as it was the case in
for the master equation defined in Eq.~(\ref{eq:master}). As we will
see in the following, and has already been extensively discussed in
\cite{Bosia2013}, this does not affect the good quality of the
approximation. Let us denote with $\vec X$ the five-dimensional vector
of components $r_1,r_2,p_1,p_2,s$ respectively. We can thus make the
following multivariate Gaussian ansatz for the probability
distribution function of $\vec X$:
\begin{equation}
\label{eq:gauss}
P^{\mathrm{Gauss}}( \vec X ) := \frac1{\sqrt{(2\pi)^5}
  \mathrm{det}C} \exp\left[-\frac{(\vec X - \vec \mu)^T 
C (\vec X - \vec \mu)}2\right]\,\,\,\,\,,
\end{equation}
which in our $5-$dimensional case depends on 20 parameters: 5 numbers
specify the mean $\vec \mu$ and 15 the covariance matrix $C$ (which is
symmetric). The key property that makes Eq.~(\ref{eq:master}) very
difficult to solve analytically is that, as shown in details in
\cite{Bosia2013}, it generates a whole hierarchy of moments such that
the lower moments are expressed in terms of higher order moments and
no moment-closure scheme can be utilized. Multivariate Gauss
distributions, on the other hand, have the useful property that all
moments can be expressed as a linear combination of just the first and
the second moments. As an illustrative example, defining $\mu_i =
E(X_i)$ and $E(X_i X_j) - E(X_i) E(X_j)$ for $i\neq j$, we could
consider the generic third moment of the distribution defined in
Eq.~(\ref{eq:gauss}) $E(X_i,X_j,X_k) = C_{ij} \mu_k + C_{ik}\mu_j +
C_{jk}\mu_i$ for $i\neq j \neq k$.

A systematic procedure to compute $\vec \mu$ and $C$ requires to
define the time dependent moment generating function:
\begin{equation}
\label{eq:momgenfun}
F_t(\vec z ) := \prod_{i=1}^5 z_i^{X_i} P_t(\vec X) 
\end{equation}
Plugging the above equation in the master equation we get the
following second order partial differential equation:
\begin{equation}
\partial_t F_t(\vec z) = {\cal H}(\vec z) F_t(\vec z) 
\end{equation} 
The moment generating function has the following properties:

\begin{eqnarray}
F({\bf z =1, q =1}) &=& 1 \; , \\ \nonumber
\partial_{z_i} F |_{\bf z =1, q =1} &=& \langle X_i\rangle \; , \\ \nonumber
\partial_{z_i}^{2} F  |_{\bf z =1, q =1} &=& \langle X_i^2\rangle - \langle X_i\rangle \; , \\ \nonumber
\partial_{z_i,z_j}^{2} F  |_{\bf z =1, q =1} &=&  \langle X_i X_j \rangle \; .
\label{property}
\end{eqnarray}
By inserting the previous definitions and imposing the Gaussian
marginalization conditions mentioned above, we obtain a system of 20
equations in 20 unknown that we can numerically solve to get the
values for $\vec \mu$ and $C$. As already shown in \cite{Bosia2013},
this approximation turns out to reproduce fairly accurately both noise
(in terms of coefficient of variation CV) of single targets and
Pearson correlation coefficient between targets, when compared with
the numerical values obtained through Gillespie algorithm.

\section*{Supplementary figures}

\begin{figure}
  \includegraphics[width=0.9\columnwidth]{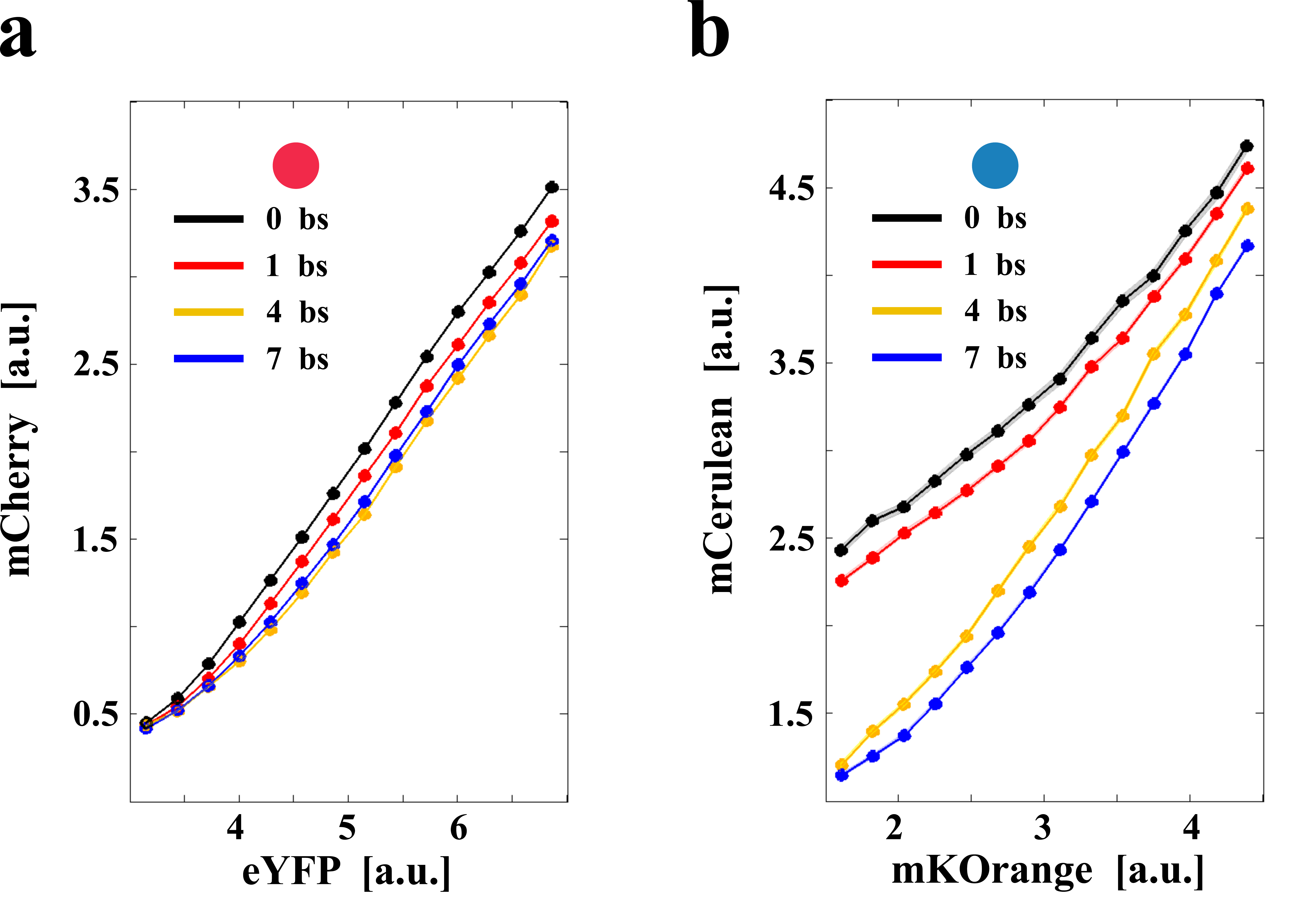}\protect\caption{\label{fig:S1}
  \textbf{Threshold for single plasmid transfections with CaCl$_{2}$}
  (a) mCherry mean fluorescence is plotted against eYFP. A threshold
  emerges when increasing mCherry bs. The same holds for mCerulean with
  respect to mKOrange (b). Shadowed strips around data are the error
  bars on the biological replicates.  Red and cyan circles in legends
  represent the mCherry and mCerulean fluorophores.}
\end{figure}

\begin{figure}
  \includegraphics[width=0.9\columnwidth]{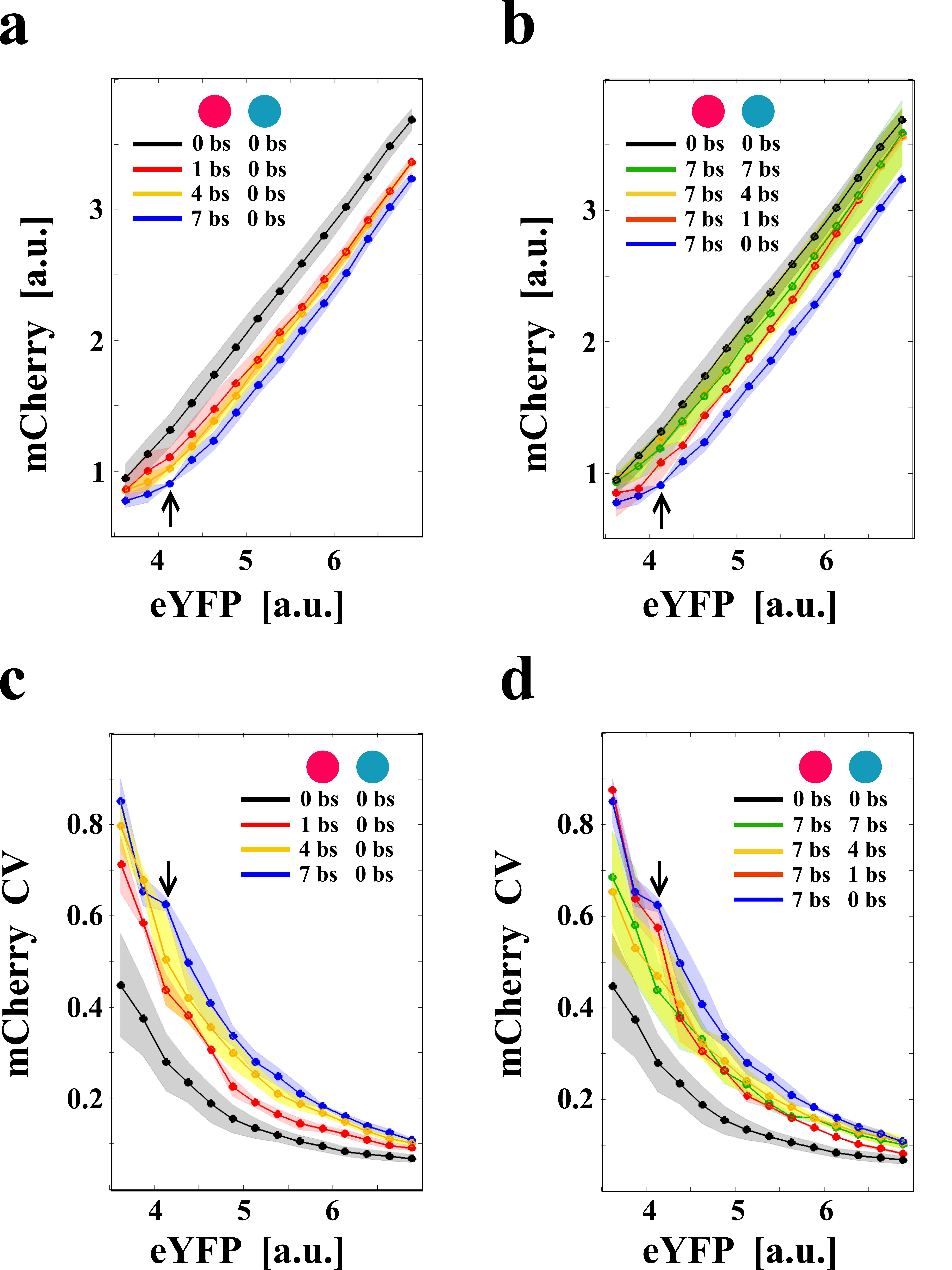}\protect\caption{\label{fig:S2}
  \textbf{Threshold and noise for CaCl$_2$ transfections}
  (a-b) mCherry mean fluorescence is plotted against eYFP. Shadowed
  strips around data are the error bars on the biological replicates. A
  threshold emerges when increasing mCherry MREs (a) while it disappears
  when increasing mCerulean MREs (b).  (c-d) mCherry coefficient of
  variation (CV) is plotted against eYFP. Shadowed strips around data
  are the error bars on the biological replicates. The CV of mCherry
  globally increases when increasing mCherry MREs (a) while decreases
  when increasing mCerulean MREs (b).  Red and cyan circles in legends
  represent the mCherry and mCerulean fluorophores. Black arrows point
  to the threshold.}
\end{figure}

\begin{figure}
  \includegraphics[width=0.9\columnwidth]{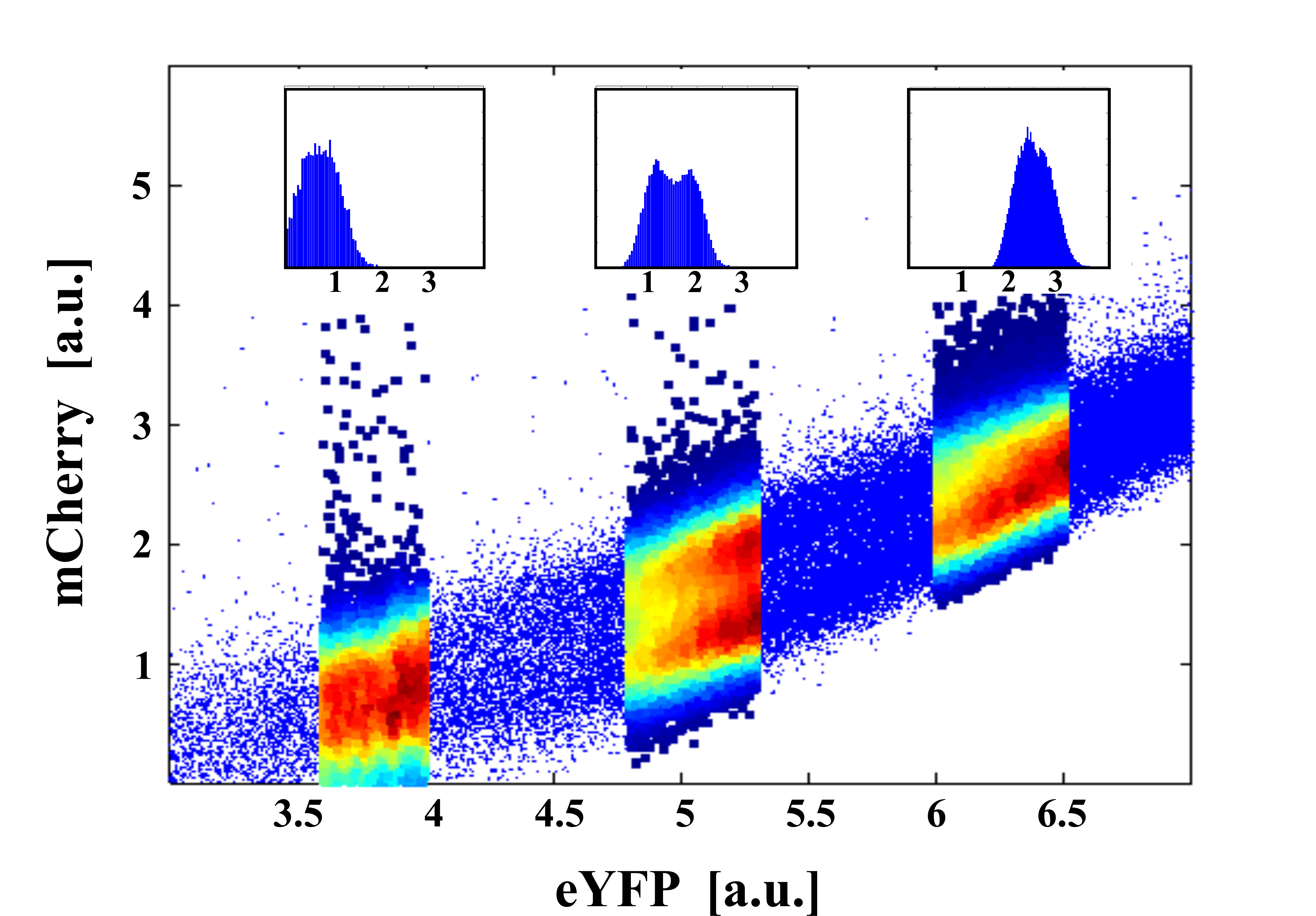}\protect\caption{\label{fig:S3}
  \textbf{Strong miRNA-target interaction engenders bimodality}
  Rough mCherry cytofluorimetry data scattered against eYFP. Soon after
  the threshold two clear phenotypes appear.}
\end{figure}

\begin{figure}
  \includegraphics[width=0.9\columnwidth]{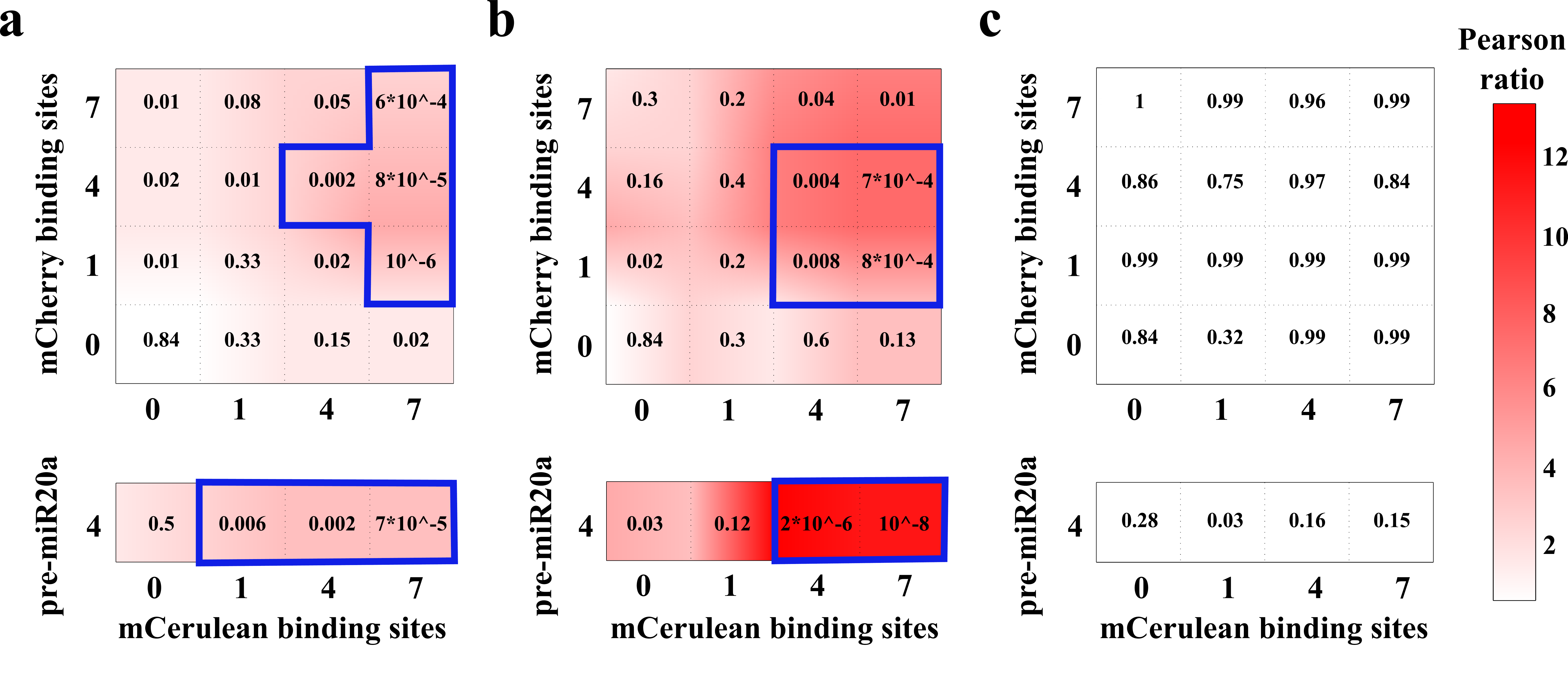}\protect\caption{\label{fig:S4}
  \textbf{Fold Pearson and p-values}
  The Pearson ratio is measured for three different values of eYFP basal
  expression: below threshold (a), around threshold (b) and above
  threshold (c). p-values are reported for each combinations of miRNA
  MREs on the two plasmids. 
  The regions inside the blue perimeters are those statistically significant with p-values 
  $<0.01$.
  As predicted by the model, the correlation is maximal around the threshold and
  could be even 12 fold higher than in the unregulated case. Blue-delimited areas
  are regions whose Pearson ratio (i.e. ratio of Pearson coefficient
  between mCherry and mCerulean possessing different MRE to the same measure in the
  absence of MRE) is statistically relevant with respect to the corresponding
  unregulated case.}
\end{figure}

\begin{figure}
  \includegraphics[width=0.9\columnwidth]{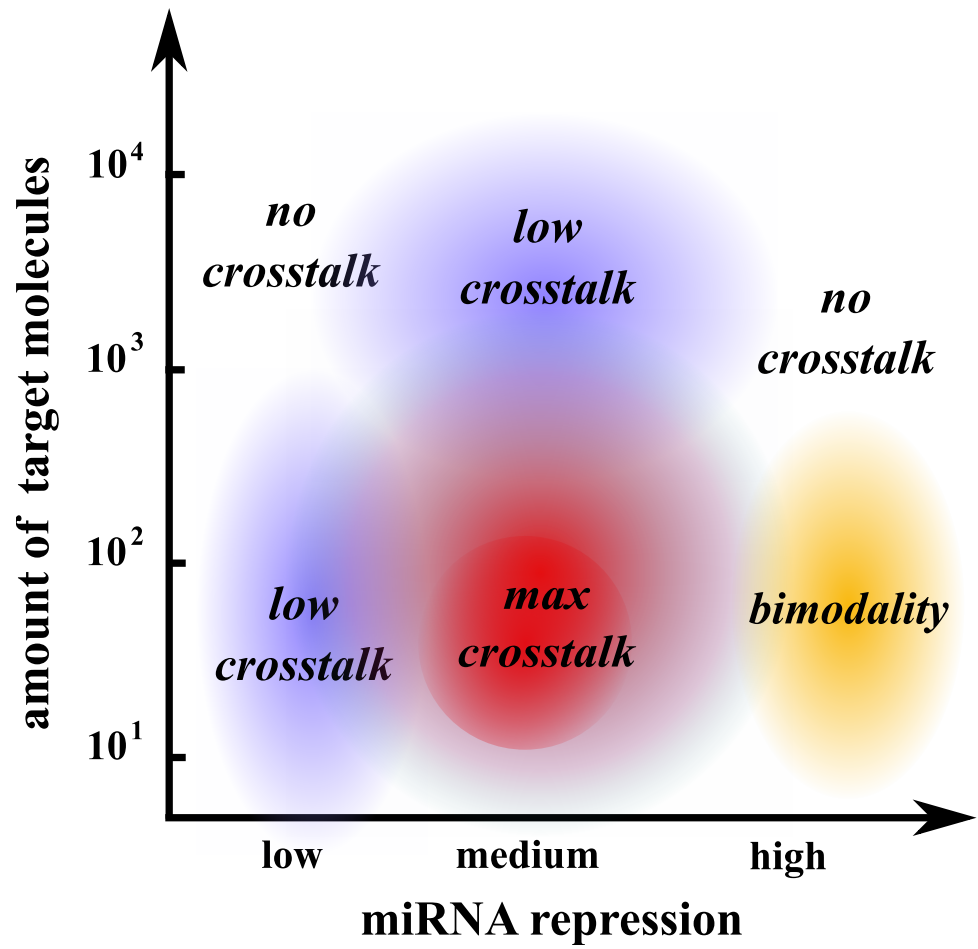}\protect\caption{\label{fig:S5}
  \textbf{Map of phenomenology}
  The cartoon shows how the crosstalk between targets and bimodality on mCherry behave varying effective miRNA respression strength and target molecule amounts.}
\end{figure}

\clearpage

\bibliographystyle{unsrt}
\bibliography{arxiv_draft}

\end{document}